\documentclass[a4paper,pre,twocolumn]{revtex4}
\usepackage{epsf}
\usepackage[dvips]{graphicx}
\usepackage{graphics}

\begin{document}
\title{Growing Surfaces with Anomalous Diffusion - \\ Results for the Fractal Kardar-Parisi-Zhang Equation}
\author{Eytan Katzav}
\email{eytak@post.tau.ac.il} \affiliation {School of Physics and
Astronomy, Raymond and Beverly Sackler Faculty of Exact Sciences,
Tel Aviv University, Tel Aviv 69978, Israel}

\begin{abstract}
In this paper I study a model for a growing surface in the
presence of anomalous diffusion, also known as the Fractal
Kardar-Parisi-Zhang equation (FKPZ). This equation includes a
fractional Laplacian that accounts for the possibility that
surface transport is caused by a hopping mechanism of a Levy
flight. It is shown that for a specific choice of parameters of
the FKPZ equation, the equation can be solved exactly in one
dimension, so that all the critical exponents, which describe the
surface that grows under FKPZ, can be derived for that case.
Afterwards, the Self-Consistent Expansion (SCE) is used to predict
the critical exponents for the FKPZ model for any choice of the
parameters and any spatial dimension. It is then verified that the
results obtained using SCE recover the exact result in one
dimension. At the end a simple picture for the behavior of the
Fractal KPZ equation is suggested, and the upper critical
dimension of this model is discussed.
\end{abstract}

\maketitle

The Kardar-Parisi-Zhang (KPZ) equation \cite{kpz86} for surface
growth under ballistic deposition was introduced as an extension
of the Edwards-Wilkinson theory \cite{barabasi95}. The interest in
the KPZ equation exceeds far beyond the interest in evolving
surfaces because of the following reasons: (a) The KPZ system is
known to be equivalent to a number of very different physical
systems. Examples are the directed polymer in a random medium,
Schrodinger equation (in imaginary time) for a particle in the
presence of a potential that is random in space and time and the
important Burgers equation from hydrodynamics \cite{barabasi95}.
(b) The second reason, that is more important, to my mind, is that
it serves as a relatively simple prototype of non-linear
stochastic field equations that are so common in condensed matter
physics.

The equation for the height of the surface at the point $\vec r$
and time $t$, $h\left( {\vec r,t} \right)$, is given by

\begin{eqnarray}
\frac{{\partial h}}{{\partial t}} = \nu \nabla ^2 h +
\frac{\lambda }{2}\left( {\nabla h} \right)^2  + \eta \left( {\vec
r,t} \right)
 \label{1},
\end{eqnarray}
where automatically the constant deposition rate is removed and
$\eta \left( {\vec r,t} \right)$ is a noise term such that
\begin{eqnarray}
\begin{array}{l}
 \left\langle {\eta \left( {\vec r,t} \right)} \right\rangle  = 0 \\
 \left\langle {\eta \left( {\vec r,t} \right)\eta \left( {\vec r',t} \right)} \right\rangle  = 2D_0 \delta \left( {\vec r - \vec r'} \right)\delta \left( {t - t'} \right) \\
 \end{array}
\label{2},
\end{eqnarray}

As can be seen in eq. (\ref{1}) the basic relaxation mechanism in
the KPZ equation is a Laplacian term that results from nearest
neighbor hopping in the growing surface. During the last years,
there has been a growing interest in other relaxational mechanisms
\cite{Metzler99}-\cite{Sokolov2001}, namely subdiffusive diffusion
that seems to appear in the context of charge transport in
amorphous semiconductors \cite{Scher75,Gu96}, NMR diffusometry in
disordered materials \cite{Klemm97}, and the dynamics of a bead in
polymer networks \cite{Amblard96}. The theoretical effort to
account for such phenomena led to the formulation of the
celebrated fractional Fokker-Planck equation (FFPE)
\cite{Metzler99}. This equation includes a fractional
(Riemann-Liouville) operator instead of the standard derivative of
the Fokker-Planck equation. Recently, Mann and Woyczynski
\cite{Mann2001} have suggested that in order to account for
experimental data, namely experiments in which impurities were
present on the growing surface \cite{Kellog94}, a modification of
the KPZ equation has to be considered. They used the observation
that the presence of an impurity can act as a strong trap for an
adatom migrating at room temperature, to conjecture that this
process corresponds to Levy flights between trap sites. This
conjecture then served as a justification for the introduction of
a fractional Laplacian into the continuum equation of the growing
surface as another relaxation mechanism. Actually, the fractional
Laplacian dominates the standard Laplacian in the KPZ equation in
the scaling regime (i.e. in the large scale limit), so that the
standard Laplacian can be ignored from the beginning. To summarize
this exposition and to be more specific, the equation they
eventually suggest to describe the growing surface in the presence
of self-similar hopping surface diffusion is the fractal KPZ
(FKPZ) equation given by
\begin{eqnarray}
\frac{{\partial h}}{{\partial t}} = \nu \left( {\Delta _\alpha  h}
\right) + \frac{\lambda }{2}\left( {\nabla h} \right)^2  + \eta
\left( {\vec r,t} \right)
 \label{3},
\end{eqnarray}
where  $\Delta _\alpha   \equiv  - \left( { - \Delta }
\right)^{{\alpha  \mathord{\left/  {\vphantom {\alpha  2}} \right.
\kern-\nulldelimiterspace} 2}}$ (or in Fourier space $\left(
\Delta _\alpha \mathcal{F} \right)\left( q \right) \equiv - \left(
{\left| q \right|^2} \right)^{{\alpha \mathord{\left/ {\vphantom
{\alpha 2}} \right. \kern-\nulldelimiterspace} 2}}
\mathcal{F}\left( q \right)$) is the fractional Laplacian, and in
our context it is more convenient to choose $\alpha  = 2 - \rho $
with $0 \le \rho < 2$ (where the special case $\rho  = 0$
corresponds to the standard KPZ equation). In addition, $\eta
\left( {\vec r,t} \right)$ is a noise characterized by
\begin{eqnarray}
\left\langle {\eta \left( {\vec r,t} \right)\eta \left( {\vec
r',t} \right)} \right\rangle  = 2D_0 \left| {\vec r - \vec r'}
\right|^{2\sigma  - d} \delta \left( {t - t'} \right)
\label{5}.
\end{eqnarray}
Actually Mann and Woyczynski \cite{Mann2001} discussed the
specific case of white noise that corresponds to $\sigma  = 0$ in
the last equation, but since the more general case does not
require special efforts we discuss the FKPZ problem with spatially
correlated noise. Furthermore, an exact solution is possible only
for a special case with correlated noise, so that the more general
discussion is also interesting on that basis.

The proposed FKPZ equation generalizes the FFPE equation mentioned
above in that it is a field equation, rather than an equation for
a single degree of freedom. Therefore, it is understood that such
a generalization is essential in order to account for the dynamics
of a whole medium experiencing anomalous diffusion, and not just
an artificial problem. Obviously, the technical mathematical
difficulties to be overcome in this nonlinear case are formidable
in comparison to those for the linear fractal kinetic equations.

However, in their paper Mann and Woyczynski \cite{Mann2001} were
not able to predict the critical exponents that describe the
surface that grows under FKPZ. But before I make any new
statements about this model let me briefly summarize the various
quantities of interest.

A very important quantity of interest is the roughness exponent
$\alpha$ that characterizes the surface in steady state. The
roughness exponent is usually defined using $W\left( {L,t}
\right)$ the roughness of the surface (that is defined as the RMS
of the height function $h\left( {\vec r,t} \right)$, in a system
of linear size L). Then, in terms of $W\left( {L,t} \right)$,
$\alpha $ is given by
\begin{eqnarray}
W\left( {L,t} \right) \propto L^\alpha
\label{7}.
\end{eqnarray}

Another important quantity of interest is the growth exponent
$\beta $ that describes the short time behavior of the roughness
$W\left( {L,t} \right)$ (with flat initial conditions)
\begin{eqnarray}
W\left( {L,t} \right) \propto t^\beta
\label{8}.
\end{eqnarray}

Finally, I introduce the dynamic exponent $z$ that describes the
typical relaxation time scale of the system (i.e. the dependence
of the equilibration time on the size of the system)
\begin{eqnarray}
t_x  \propto L^z
\label{9}.
\end{eqnarray}

It is well known \cite{barabasi95} that these three exponents are
not independent and that under very general considerations one
should expect the following scaling relation
\begin{eqnarray}
z = {\alpha  \mathord{\left/ {\vphantom {\alpha  \beta }} \right.
\kern-\nulldelimiterspace} \beta }
\label{10}
\end{eqnarray}
(this relation is a direct consequence of the Family-Vicsek
scaling relation \cite{Family85}.

In addition, for the KPZ equation there is another scaling
relation, that comes from a symmetry of the equation under
infinitesimal tilting of the surface (this symmetry is just the
famous Galilean invariance of the Burgers equation)
\cite{barabasi95}
\begin{equation}
\alpha+z=2
\label{10.5}.
\end{equation}
It can easily be checked that this symmetry holds in the case of
FKPZ as well, because the fractal Burgers equation is evidently
invariant under Galilean transformation (see ref. \cite{Mann2001},
eq. (7.8)). Therefore, the last scaling relation is also relevant
in the this discussion. Hence both scaling relations, reduce the
number of unknown exponents to just one (out of the three we
started with). In some cases an extra scaling relation is
possible (for example, in the case of the KPZ equation with
long-range noise - see refs. \cite{Frey99, Janssen99, medina89}),
so that the exponents can be obtained exactly by power-counting.
Employing the terminology of the Dynamic Renormalization Group
(DRG) approach this can be explained by saying that certain terms
in the dynamic action do not renormalize, and so an extra scaling
relation arises (see \cite{Frey94}). This kind of solutions are
naturally available also in the present problem. Whenever the
exponents are obtained due to such an extra condition it will be
specifically pointed out.

In this paper I show that for a specific choice of parameters of
the FKPZ equation (namely $\rho  = 2\sigma$), the equation can be
solved exactly in one dimension, so that all the critical
exponents can be derived easily for that case. Afterwards, in
order to give a more complete picture (i.e. for any dimension d,
and any spatial correlation index $\sigma$) I apply a method
developed by Schwartz and Edwards
\cite{schwartz92}-\cite{katzav99} (also known as the
Self-Consistent-Expansion (SCE) approach). This method has been
previously applied successfully to the KPZ equation. The method
gained much credit by being able to give a sensible prediction for
the KPZ critical exponents in the strong coupling phase, where
many Renormalization-Group (RG) approaches failed, as well as
Dynamic Renormalization Group (DRG) \cite{kpz86, medina89}
(actually, it can be shown that the strong coupling regime is
inaccessible by DRG even when it is used to all orders
\cite{Lassig95, Wiese98}). It is then verified that the results
obtained using SCE recover the exact result in one dimension.

As mentioned above, for the specific case $\rho  = 2\sigma$ in one
dimension an exact solution can be found for the FKPZ problem
using the Fokker-Planck equation associated with its Langevin form
(i.e. eq. (\ref{3})). This particular choice of parameters
corresponds to a situation where the fractional exponent $\rho$
equals the exponent $2\sigma$ that describes the decay of spatial
correlations in the noise. Since kind this exact solutions is
familiar in the KPZ community I will simply state the final
results given by $\alpha=1/2$ and $z=3/2$ (using the scaling
relation (\ref{10.5})). Notice that these critical exponents
extend the classical one dimensional KPZ exponents for non zero
$\rho$'s and $\sigma$'s, as the classical KPZ case corresponds to
$\rho = 2\sigma = 0$.

Now, the Self-Consistent Expansion (SCE) is applied in order to
learn about the behavior of this system in more general contexts
(other dimensions and cases where $\rho  \ne 2\sigma $). SCE's
starting point is the Fokker-Planck form of FKPZ, from which it
constructs a self-consistent expansion of the distribution of the
field concerned.

The expansion is formulated in terms of $\phi _q $ and $\omega _q
$, where $\phi _q$ is the two-point function in momentum space,
defined by $\phi _q  = \left\langle {h_q h_{ - q} } \right\rangle
_S $, (the subscript S denotes steady state averaging), and
$\omega _q $ is the characteristic frequency associated with
$h_q$. It is expected that for small enough $q$, $\phi _q $ and
$\omega _q $ are power laws in $q$,

\begin{eqnarray}
\phi _q  = Aq^{ - \Gamma } \qquad {\rm{ and }} \qquad \omega _q  =
Bq^z
\label{16}
\end{eqnarray}
where $z$ is just the dynamic exponent, and the exponent $\Gamma$
is related to the roughness exponent $\alpha $ by

\begin{equation}
 \alpha  = \frac{{\Gamma  - d}}{2}
\label{18}.
\end{equation}

The main idea is to write the Fokker-Planck equation ${{\partial
P} \mathord{\left/  {\vphantom {{\partial P} {\partial t}}}
\right. \kern-\nulldelimiterspace} {\partial t}} = OP$ in the
form ${{\partial P} \mathord{\left/  {\vphantom {{\partial P}
{\partial t}}} \right. \kern-\nulldelimiterspace} {\partial t}} =
\left[ {O_0  + O_1  + O_2 } \right]P$, where $O_0$ is to be
considered zero order in some parameter$\lambda$, $O_1$ is first
order and $O_2$ is second order. The evolution operator $O_0$ is
chosen to have a simple form $O_0  =  - \sum\limits_q
{\frac{\partial }{{\partial h_q }}\left[ {D_q \frac{\partial
}{{\partial h_{ - q} }} + \omega _q h_q } \right]} $, where
${{D_q } \mathord{\left/  {\vphantom {{D_q } {\omega _q }}}
\right.  \kern-\nulldelimiterspace} {\omega _q }} = \phi _q$.
Note that at present $\phi _q$ and $\omega _q$ are not known. I
obtain next an equation for the two-point function. The expansion
has the form $\phi _q  = \phi _q  + c_q \left( {\left\{ {\phi _p
} \right\},\left\{ {\omega _p } \right\}} \right)$, because the
lowest order in the expansion already yields the unknown $\phi
_q$. In the same way an expansion for $\omega _q$ is also
obtained in the form $\omega _q  = \omega _q + d_q \left(
{\left\{ {\phi _p } \right\},\left\{ {\omega _p } \right\}}
\right)$. Now, the two-point function and the characteristic
frequency are thus determined by the two coupled equations

\begin{eqnarray}
c_q \left( {\left\{ {\phi _p } \right\},\left\{ {\omega _p }
\right\}} \right) = 0  \qquad {\rm{ and }} \qquad d_q \left(
{\left\{ {\phi _p } \right\},\left\{ {\omega _p } \right\}}
\right) = 0
\label{19}.
\end{eqnarray}

These equations can be solved exactly in the asymptotic limit to
yield the required scaling exponents governing the steady state
behavior and the time evolution. Working to second order in the
expansion, one gets the two coupled integral equations

\begin{eqnarray}
D_q  &-& \nu _q \phi _q  + 2\sum\limits_{\ell ,m}
{\frac{{\left|M_{q\ell m}\right|^2 \phi _\ell  \phi _m }}{{\omega
_q + \omega _\ell + \omega _m }}} - 2\sum\limits_{\ell ,m}
{\frac{{M_{q\ell m} M_{\ell mq} \phi _m \phi _q }}{{\omega _q  +
\omega _\ell   + \omega _m }}} \nonumber \\
&-& 2\sum\limits_{\ell ,m} {\frac{{M_{q\ell m} M_{m\ell q} \phi
_\ell  \phi _q }}{{\omega _q + \omega _\ell   + \omega _m }}}  = 0
\label{20},
\end{eqnarray}
and
\begin{eqnarray}
\nu _q  - \omega _q  - 2\sum\limits_{\ell ,m} {M_{q\ell m}
\frac{{M_{\ell mq} \phi _m  + M_{m\ell q} \phi _\ell  }}{{\omega
_\ell   + \omega _m  + \omega _q }}}  = 0
\label{21},
\end{eqnarray}
where $D_q=D_0 q^{-2\sigma}$, $\nu _q  = \nu q^{2 - \rho }$ and
$M_{q\ell m} = \frac{\lambda }{{2\sqrt \Omega  }}\left( {\vec \ell
\cdot \vec m} \right)\delta _{q,\ell  + m} $. In addition, in
deriving eq. (\ref{21}) I have used the Herring consistency
equation \cite{Herring}. In fact Herring's definition of $\omega
_q $ is one of many possibilities, each leading to a different
consistency equation. But it can be shown, as previously done in
\cite{schwartz98}, that this does not affect the exponents
(universality).

A detailed solution of equations (\ref{20}) and (\ref{21}) in the
limit of small $q's$ (i.e. large scales) in the line of refs.
\cite{schwartz98}-\cite{katzav99} yields a rich family of
solutions that I shall describe immediately.

First, there are two kinds of weak-coupling solutions - both with
a dynamic exponent $z  = 2 - \rho $ (they are called weak-coupling
because they are exactly the solutions obtained in the case of the
Fractal Edwards- Wilkinson equation, see \cite{Mann2001}). Now,
when the spatial correlations of the noise are relevant (i.e.,
when $\sigma  > 0$) I obtain the solution $\Gamma  = 2 - \rho  +
2\sigma $- provided that $d > 2 - 3\rho  + 2\sigma $. But if the
spatial correlations of the noise are not relevant (i.e., when
$\sigma  \le 0$) I obtain the simpler solution $\Gamma  = 2 -
\rho$ - provided that $d > 2 - 3\rho$.

The second type of solutions is strong coupling solutions that
obey the well known scaling relation $z  = \frac{{d + 4 - \Gamma
}}{2}$ obtained from eq. (\ref{21}) (this scaling relation is just
the above-mentioned scaling relation $\alpha+z=2$ that is
naturally obeyed by our analysis). The first strong coupling
solution is determined by the combination of the scaling relation
and the transcendental equation $F\left( {\Gamma ,z } \right) =
0$, where F is given by

\begin{eqnarray}
 F\left( {\Gamma ,z } \right) &=&  -  \int {d^d t\frac{{\vec t \cdot \left( {\hat e - \vec t} \right)}}{{t^z   + \left| {\hat e - \vec t} \right|^z
 +1}}\times}  \nonumber \\
&\times&  \left[ {\left( {\hat e \cdot \vec t} \right) t^{ - \Gamma }  + \hat e \cdot \left( {\hat e - \vec t} \right) \left| {\hat e - \vec t} \right|^{ - \Gamma } } \right]  \nonumber\\
&+& \int {d^d t\frac{{\left[ {\vec t \cdot \left( {\hat e - \vec
t} \right)} \right]^2 }}{{t^z   + \left| {\hat e - \vec t}
\right|^z   + 1}}t^{ - \Gamma } \left| {\hat e - \vec t} \right|^{
- \Gamma } } \label{22},
\end{eqnarray}
and $\hat e$ is a unit vector in an arbitrary direction. \\
This solution is valid as long as the solutions of the last
equations satisfy the following condition $\Gamma  > \max \left\{
{\frac{{d + 4 + 4\sigma }}{3},\frac{{d + 4}}{3},d + 2\rho }
\right\}$.

It turns out that for $d = 1$ the equation $F\left( {\Gamma ,z
\left( \Gamma  \right)} \right) = 0$ is exactly solvable, and
yields $\Gamma  = 2$ and $z  = \frac{3}{2}$ (it can be checked
immediately by direct substitution). In this case the validity
condition reads $2 > \max \left\{ {\frac{{5 + 4\sigma }}{3},1 +
2\rho } \right\}$ or equivalently $0 < \rho  < {1 \mathord{\left/
{\vphantom {1 2}} \right. \kern-\nulldelimiterspace} 2}$ and
$\sigma  < {1 \mathord{\left/ {\vphantom {1 4}} \right.
\kern-\nulldelimiterspace} 4}$. By using eq. (\ref{18}) I
translate the results into $\alpha  = \frac{1}{2}$ and $z =
\frac{3}{2}$ that are precisely the exact result presented above.

It should also be mentioned that for $d \ge 2$ such an exact
solution in closed form cannot be found, and one has to solve
numerically the equation $F\left( {\Gamma ,z \left( \Gamma
\right)} \right) = 0$. For convenience I denote the numerical
value of this solution by $\Gamma _0 \left( d \right)$. For
example, in two dimensions I obtain $\Gamma _0 \left( 2 \right) =
2.59$.

The second strong coupling solution is obtained by power-counting,
and it is relevant when $\sigma  > 0$ (i.e. when the spatial
correlations of the noise are relevant), $d < 2 + 2\sigma $ and
$F\left( {\Gamma ,z \left( \Gamma  \right)} \right) < 0$ (the last
condition turns out to be equivalent to the condition $\Gamma  >
\Gamma _0 \left( d \right)$ because from this value of $\Gamma $
and on the function $F\left( {\Gamma ,z \left( \Gamma  \right)}
\right)$ is negative). This power counting solution can be written
in closed form and is given by $z  = \frac{{d + 4 - 2\sigma }}{3}$
and $\Gamma  = \frac{{d + 4 + 4\sigma }}{3} = z  + 2\sigma $.

The third strong coupling solution (that is in some sense the only
"genuine" FKPZ solution, in the sense that it is the only solution
that is dramatically influenced by the fractional Laplacian, and
at the same time is not a solution of the Fractal
Edwards-Wilkinson equation) is also a obtained by power counting.
More specifically, it is determined by the combination of the
scaling relation $z = \frac{{d + 4 - \Gamma }}{2}$ and the extra
relation $d + 4 - 2\Gamma  - z  = 2 - \rho - \Gamma $. This
solution can be written in closed form as $\Gamma = d + 2\rho$ and
$z  = 2 - \rho $. It turns out that this phase is relevant when $d
> \max \left\{ {2 - 3\rho ,2 - 3\rho  + 2\sigma } \right\}$. In
addition, it is needed that $F\left( {\Gamma ,z \left( \Gamma
\right)} \right)$ is positive. Therefore, this solution is
possible only when $\Gamma  < \Gamma _0 \left( d \right)$.

The following table summarizes all the possible phases found in
this paper

\begin{table}[ht]
\begin{tabular}{|c|c|l|}
\hline $z$ & $\Gamma$ & $Validity$ \\
\hline\hline $2-\rho$ & $2 - \rho  + 2\sigma $ & $\sigma  > 0$ and $d > 2  - 3\rho + 2\sigma$ \\

$2-\rho$ & $2-\rho$ & $\sigma  \le 0 $ and $d > 2
- 3\rho $ \\

$2-\rho$ & $d + 2\rho$ & $0 < \rho  < 1$ and \\
&&$d > \max \left\{ {2 - 3\rho ,2 - 3\rho  + 2\sigma } \right\}$\\

$\frac{{d + 4 -  2\sigma }}{3}$ &  $\frac{{d + 4 + 4\sigma }}{3}$
& $\sigma  > 0 $, $d < 2 + 2\sigma$ and \\
&&$d + 4 + 4\sigma  > 3\Gamma _0 \left( d \right)$\\

$\frac{{d + 4 - \Gamma _0 \left( {d} \right)}}{2}$ & $\Gamma _0
\left( {d} \right)$ & $\Gamma_0 \left( {d} \right) > \max \left\{
{\frac{{d + 4 +
4\sigma }}{3},\frac{{d + 4}}{3},d + 2\rho } \right\}$ \\
\hline
\end{tabular}
\caption{A complete description of all the possible phases of the
FKPZ problem, for any value of $d,\rho $ and $\sigma $. The first
two columns give the scaling exponents $z $ and $\Gamma $ for a
particular phase, and the third column states each phase's
validity condition. Note that $\Gamma _0 \left( {d} \right)$ is
the numerical solution of the transcendental equation $F\left(
{\Gamma ,z} \right) = 0$ with the scaling relation $z = \frac{{d +
4 - \Gamma }}{2}$ - if such a solution exists.} \label{exp_vals}
\end{table}

In order to gain more insight on this system, it might be
interesting to specialize to two extreme cases: namely $\rho=0$
and $\sigma \neq 0$ vs. $\rho \neq 0$ and $\sigma = 0$. The first
case (namely $\rho=0$ and $\sigma \neq 0$) corresponds to the
local KPZ problem with long-range noise. This problem has been
studied in the past using various methods - for example: DRG
\cite{Frey99, Janssen99, medina89}, Mode-Coupling \cite{chat98}
and SCE \cite{katzav99}. All methods agree on the basic picture
that for a big enough noise exponent ($\sigma$) one obtains a
power-counting strong-coupling solution, given by $z=\frac{{d + 4
- 2\sigma }}{3}$ ($z$ is the dynamic exponent). The controversy
between the different methods is over the values of the scaling
exponents for smaller values of $\sigma$, and on the critical
value $\sigma_0$ that separates between the two phases. Not
surprisingly, the results given here agree with the previous SCE
result presented in ref. \cite{katzav99}.

The second case (namely $\rho \neq 0$ and $\sigma = 0$)
corresponds to the Fractal KPZ problem with white noise that is
the original equation suggested by Mann and Woyczynski
\cite{Mann2001}. Specializing the general picture of phases
presented in Table I above to the case $\sigma = 0$ yields new
phase diagrams, namely a separate phase diagram for every
dimension $d$. For example, in two dimensions there are three
possible phases: I. The standard KPZ phase given by $\Gamma _0
\left( {2} \right)=2.59$ that is valid for $0 \le \rho \le 0.295$.
II. The weak-coupling phase given by $\Gamma = 2-\rho$ that is
valid for any $\rho > 0$. III. The "third" strong coupling
solution given by $\Gamma = d+2\rho$ that is possible for $0 <
\rho \le 1$. In the first phase the dynamic exponent is
$z=1.705$, while the other two phases share the same dynamic
exponent of $z=2-\rho$. The possible phases in two dimensions are
presented in Fig. 1.

\begin{figure}[htb]
\includegraphics[width=6cm]{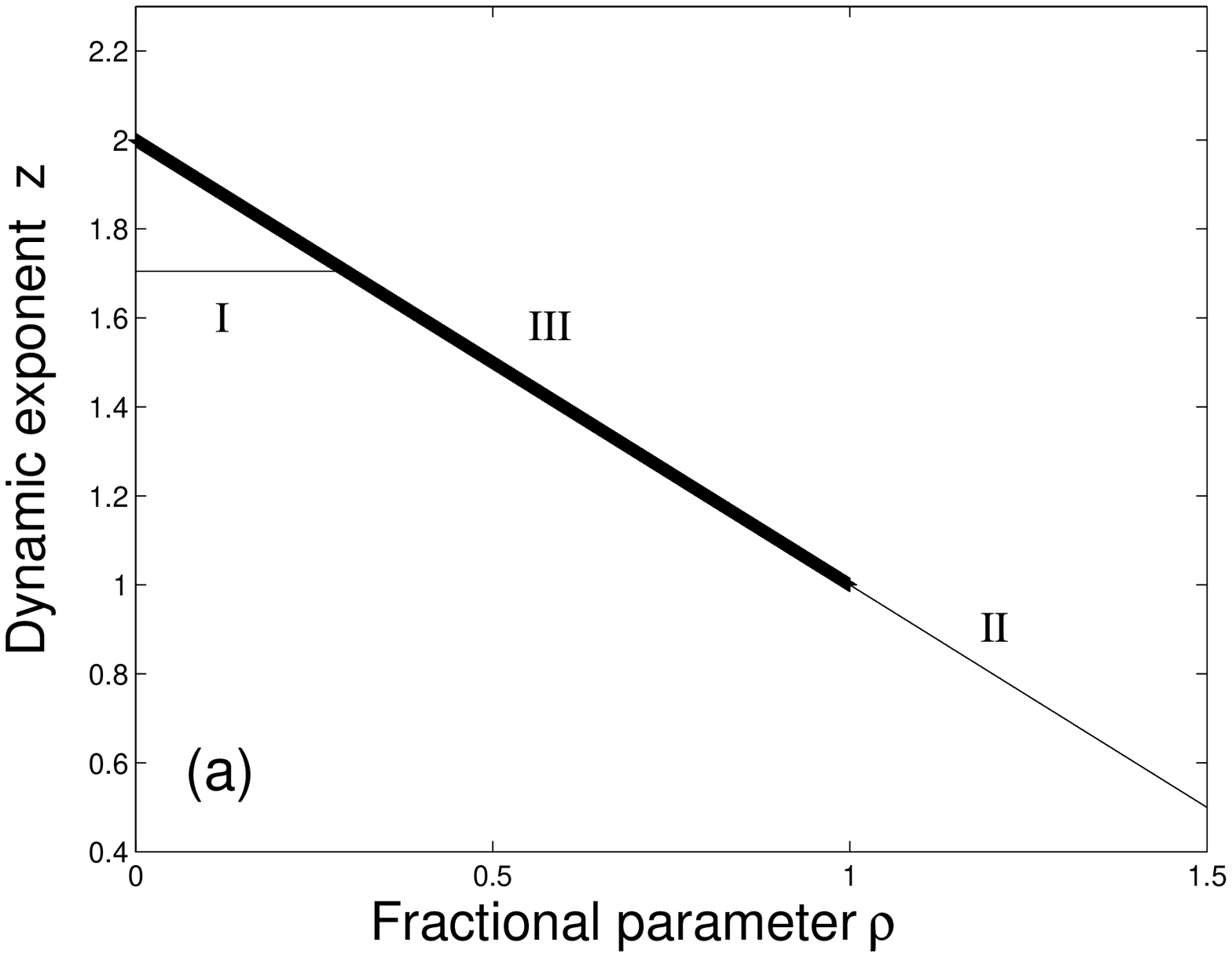}
\includegraphics[width=6cm]{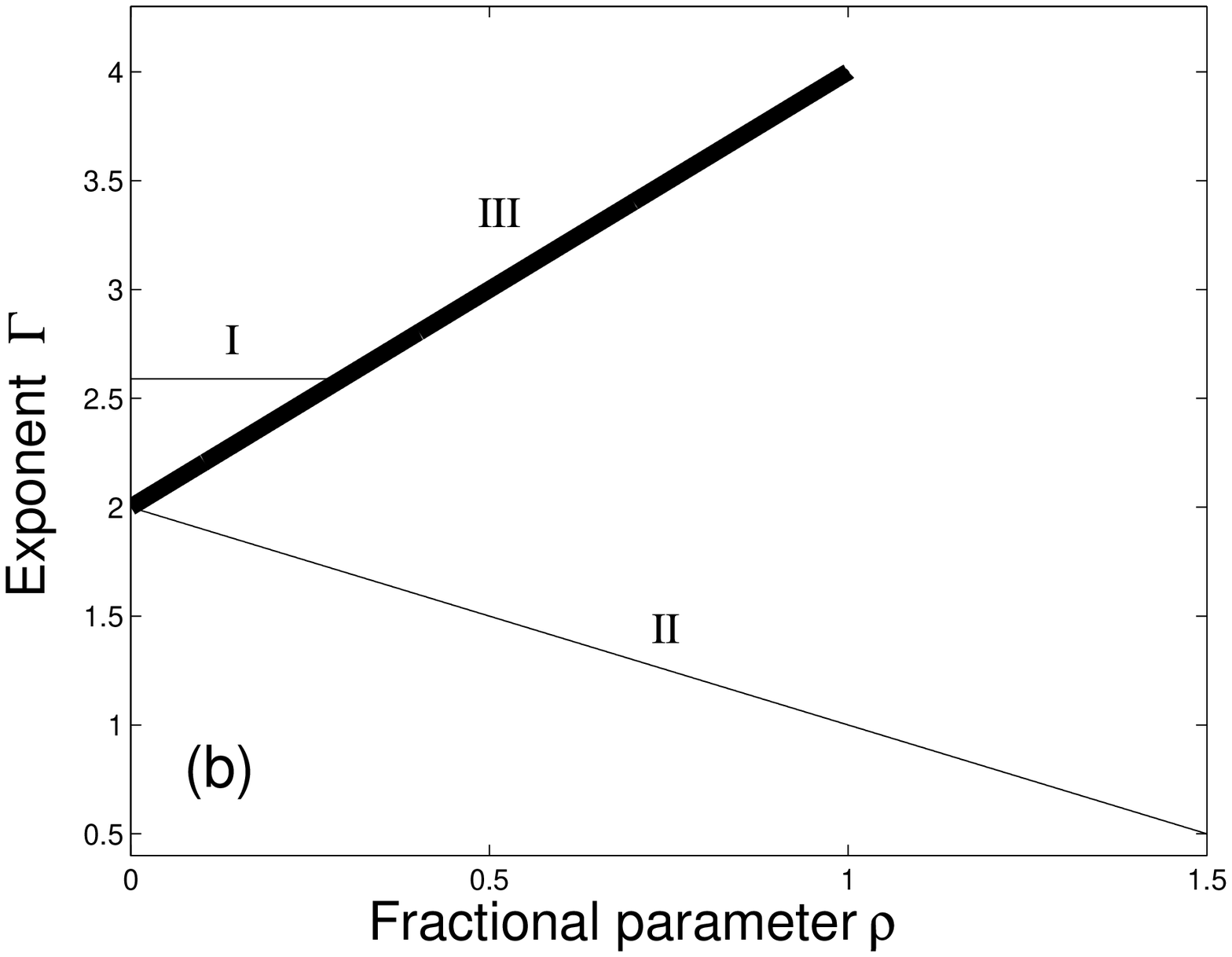}
\caption{The values of the scaling exponents (a) $z$ and (b)
$\Gamma$ as a function of the fractional parameter $\rho $, for
uncorrelated noise ($\sigma=0$) in two-dimensions.} \label{exps2}
\end{figure}

The described two-dimensional (and white-noise) scenario is quite
typical and appears in other dimensions as well - namely there
are usually three possible phases that are possible for different
values of $\sigma$ with possible phase transitions between them
(as in the usual KPZ scenario - the phase transition is
controlled by the strength of the dimensionless coupling
constant). More precisely, this picture extends up to the upper
critical dimension of the original KPZ problem ($d_{uc}^{KPZ}$)
where the first phase (phase I in Fig. 1) disappears (see the
discussion below).

At this point, the full picture of possible solutions might seem
too complex, so I want to suggest the following simple
interpretation for the behavior of the Fractal KPZ equation. If
you remember, the starting point of this discussion was the
introduction of the fractional Laplacian into the KPZ equation.
This immediately implies that faster relaxations are now possible
(faster in the sense that a smaller dynamical exponent is
expected) when compared with the Edwards-Wilkinson equation.
However, it is well known that already the KPZ nonlinearity
introduces faster relaxations (at least for dimensions lower than
the upper critical dimension). Therefore, in the FKPZ system, the
dynamics is controlled by the fastest "component": if the
dynamical exponent of the classical KPZ system is smaller than $2
- \rho $ then it dominates, otherwise the new fractional
dynamical exponent $z = 2 - \rho $ controls the dynamics.
However, at this point the picture gets a little bit more
complicated (just like in the classical KPZ case), namely there
are several possible phases with this new fractional dynamical
exponent $z = 2 - \rho $, and the transition between these phases
is controlled by the strength of dimensionless coupling constant.

These conclusions have an important implication for the upper
critical dimension of the FKPZ model (i.e. the dimension above
which the dynamical exponent is the same as that of the linear
theory). Namely, it is turns out that the FKPZ equation always
has an upper critical dimension that is determined by the
relation $\Gamma _0 \left( {d_{uc}^{FKPZ} } \right) =
d_{uc}^{FKPZ}  + 2\rho $ (or alternatively, using the roughness
exponent $\alpha $ - the upper critical dimension is the
dimension where $\alpha _{KPZ} \left( {d_{uc}^{FKPZ} } \right) =
\rho $). Notice, that this result does not dependent on the
ongoing debate over the existence of the upper critical dimension
for the KPZ system (see \cite{UCD98}-\cite{Marinari2002}).
Actually, it merely requires that the roughness exponent of the
classical KPZ system becomes arbitrarily small in higher
dimensions - an assumption that is generally accepted (only Tu
\cite{Tu94} had a different conjecture).

As one can see, the results obtained using the Self-Consistent
Expansion (SCE) are quite general, and cover all possible values
of the relevant parameters ($\rho $ and $\sigma $) as well as
dimensions. It is also easily verified that these results recover
the exact result obtained at the beginning of this paper for the
case $\rho  = 2\sigma $. This situation, suggest that the SCE
method is generally appropriate when dealing with non-linear
continuum equations. \\

Acknowledgement: I would like to thank Moshe Schwartz for useful
discussions.


\newpage


\begin{thebibliography}{34}

\bibitem{kpz86} M. Kardar, G. Parisi and Y.-C. Zhang,
  {\it Phys. Rev. Lett. {\bf 56}}, 889 (1986).
\bibitem{barabasi95} A.-L. Barabasi and H. E. Stanley, Fractal Concepts in Surface Growth (Cambridge Univ. Press,
Cambridge, 1995).
\bibitem{Mann2001} J.A. Mann Jr. and W.A. Woyczynski,
 {\it Physica A {\bf 291}}, 159 (2001).
\bibitem{Kellog94} G.L. Kellog,
  {\it Phys. Rev. Lett. {\bf 72}}, 1662 (1994).
\bibitem{Family85} F. Family and T. Vicsek,
  {\it J. Phys. A {\bf 18}}, L75 (1985).
\bibitem{schwartz92} M. Schwartz and S.F. Edwards,
 {\it Europhys. Lett. {\bf 20}}, 301 (1992).
\bibitem{schwartz98} M. Schwartz and S.F. Edwards,
 {\it Phys. Rev. E {\bf 57}}, 5730 (1998).
\bibitem{katzav99} E. Katzav and M. Schwartz,
 {\it Phys. Rev. E {\bf 60}}, 5677 (1999).
\bibitem{medina89} E. Medina, T. Hwa, M. Kardar and Y.C. Zhang,
  {\it Phys. Rev. A {\bf 39}}, 3053 (1989).
\bibitem{Herring} J.R. Herring,
 {\it Phys. Fluids {\bf 8}}, 2219 (1965); {\bf 9}, 2106 (1966).
\bibitem{UCD98} T. Ala-Nissila,
  {\it Phys. Rev. Lett. {\bf 80}}, 887 (1998).
  \\ J. M. Kim, {\it ibid. {\bf 80}}, 888 (1998).
  \\ M. Lassig and H. Kinzelbach, {\it ibid. {\bf 80}}, 889 (1998).
\bibitem{Perlsman96} E. Perlsman and M. Schwartz,
 {\it Physica A {\bf 234}}, 523 (1996).
\bibitem{Castellano98} C. Castellano, M. Marsili and L. Pietronero,
  {\it Phys. Rev. Lett. {\bf 80}}, 3527 (1998).
  \\ C. Castellano, A. Gabrielli, M. Marsili, M.A. Munoz and L. Pietronero,
  {\it Phys. Rev. E {\bf 58}}, 5209 (1998).
\bibitem{Halpin90} T. Halpin-Healy,
  {\it Phys. Rev. A {\bf 42}}, 711 (1990).
\bibitem{Blum95} T. Blum and A. J. McKane,
  {\it Phys. Rev. E {\bf 52}}, 4741 (1995).
\bibitem{Bouchaud93} J-P. Bouchaud and M. E. Cates,
  {\it Phys. Rev. E {\bf 47}}, 1455 (1993).
  \\ erratum), {Phys. Rev. E {\bf 48}}, 653 (1993).
\bibitem{Katzav2002} E. Katzav and M. Schwartz,
 {\it Physica A {\bf 309}}, 69 (2002).
\bibitem{Cook90} J. Cook and B. Derrida,
 {\it J. Phys. A {\bf 23}}, 1523 (1990).
\bibitem{Colaiori2001} F. Colaiori and M. A. Moore,
  {\it Phys. Rev. Lett. {\bf 86}}, 3946 (2001).
\bibitem{Marinari2002} E. Marinari, A. Pagnani, G. Parisi and Z. R\'{a}cz,
  {\it Phys. Rev. E {\bf 65}}, 26136 (2002).
\bibitem{Tu94} Y. Tu,
  {\it Phys. Rev. Lett. {\bf 73}}, 3109 (1994).
\bibitem{Metzler99} R. Metzler, E.Barkai and J. Klafter,
  {\it Phys. Rev. Lett. {\bf 82}}, 3563 (1999).
\bibitem{Metzler2000} R. Metzler and J. Klafter,
  {\it Chem. Phys. Lett. {\bf 321}}, 238 (2000).
\bibitem{Granek2001} R. Granek and J. Klafter,
  {\it Europhys. Lett. {\bf 56}}, 15 (2001).
\bibitem{Sokolov2001} M. Sokolov, J. Klafter and A. Blumen,
  {\it Phys. Rev. E {\bf 64}}, 21107 (2001).
\bibitem{Scher75} H. Scher and E. Montroll,
  {\it Phys. Rev. B {\bf 12}}, 2455 (1975).
\bibitem{Gu96} Q. Gu, E. A. Schiff, S. Grebner and R. Schwartz,
  {\it Phys. Rev. Lett. {\bf 76}}, 3196 (1996).
\bibitem{Klemm97} A. Klemm, H.-P. M\"{u}ller and R. Kimmich,
  {\it Phys. Rev. E {\bf 55}}, 4413 (1997).
\bibitem{Amblard96} F. Amblard, A. C. Maggs, B. Yurke, A. N. Pargellis and S. Leibler,
  {\it Phys. Rev. Lett. {\bf 77}}, 4470 (1996).

\bibitem{Lassig95} M. L\"assig,  {\it Nucl. Phys. B {\bf 448}}, 559 (1995).
\bibitem{Wiese98} K.J. Wiese,  {\it J. Stat. Phys. {\bf 93}}, 143 (1998).

\bibitem{Frey99} E. Frey, U.C. T\"auber and H.K. Janssen,
  {\it Europhys. Lett. {\bf 47}}, 14 (1999).
\bibitem{Janssen99} H.K. Janssen, U.C. T\"auber and E. Frey,
  {\it Europhys. J. B {\bf 9}}, 491 (1999).
\bibitem{chat98} A. Kr. Chattapohadhyay and J. K. Bhattacharjee,
 {\it Europhys. Lett. {\bf 42}}, 119 (1998).

\bibitem{Frey94} E. Frey and U.C. T\"auber,
  {\it Phys. Rev. E {\bf 50}}, 1024 (1994).

\end{thebibliography}
\end{document}